# Supercontinuum generation in media with sign-alternated dispersion

HAIDER ZIA[1*], NIKLAS M. LÜPKEN[2], TIM HELLWIG [2], CARSTEN FALLNICH[2,1], KLAUS-J. BOLLER[1,2]

[1] *University of Twente, Department Science & Technology, Laser Physics and Nonlinear Optics Group, MESA+ Research Institute for Nanotechnology, Enschede 7500 AE, The Netherlands*
[2] *University of Münster, Institute of Applied Physics, Corrensstraße 2, 48149 Münster, Germany*
**h.zia@utwente.nl*

Abstract:

When an ultrafast optical pulse with high intensity is propagating through transparent material a supercontinuum can be coherently generated by self-phase modulation, which is essential to many photonic applications in fibers and integrated waveguides. However, the presence of dispersion causes stagnation of spectral broadening past a certain propagation length, requiring an increased input peak power for further broadening. Overcoming such spectral stagnation will be key to achieve practical integrated supercontinuum devices. We present a concept to drive supercontinuum generation with significantly lower input power by counteracting spectral stagnation via repeatedly alternating the sign of group velocity dispersion along the propagation. We demonstrate the effect experimentally in dispersion alternating fiber in excellent agreement with modeling, revealing almost an order of magnitude reduced peak power compared to uniform dispersion. Calculations reveal a similar power reduction also with integrated optical waveguides, simultaneously with a significant increase of flat bandwidth, which is important for on-chip broadband photonics.

**Introduction**

Supercontinuum generation (SCG) is a nonlinear optical process where injecting intense, ultrashort pulses in transparent optical materials generates light with ultra-wide spectral bandwidth[1–3]. Maximum coherence is achieved when spectral broadening is dominated by self-phase modulation based on the intensity dependent Kerr index[4]. The achieved bandwidths, often spanning more than two octaves[5–11] have become instrumental in a plethora of fields[12]. Examples are metrology based on optical frequency combs[13–17], optical coherence tomography[18–21], ranging[22], or sub-cycle pulse compression[23–25].

The widest spectral bandwidths at lowest input power are typically achieved in waveguiding geometries, such as in fibers[26–28] and integrated optical waveguides on a chip[10,11,29,30] because transverse confinement of light enhances the intensity, which promotes the nonlinear generation of additional frequency components. However, although significant spectral broadening can be achieved with low input power using highly nonlinear materials[31], generating bandwidth beyond one or two octaves still requires significant input peak powers in the order of several kilowatt[11,32,33], hundreds of kilowatt [34], or close to megawatt levels[3,35]. The reason is that dispersion terminates spectral broadening beyond a certain propagation length, which means that the peak power required for a desired bandwidth cannot be reduced by extending the propagation length.

Such lack of scalability towards lower input power constitutes a critical limitation for implementations of supercontinuum generation, specifically, in all-integrated optical systems, for the generation of high repetition rate supercontinuum pulses and for supercontinuum generation with long input pulse durations. In case of normal dispersion (positive group velocity dispersion) the reason for termination of broadening is that initial pulse duration gets stretched and the waveguide-internal peak power is lowered vs propagation. This brings self-phase modulation, the main mechanism of coherent spectral broadening[36], to stagnation past a certain propagation length[4,36,37].

Lower input peak powers are sufficient with anomalous dispersion because negative group velocity dispersion leads to pulse compression by making use of the additionally generated bandwidth. However, spectral stagnation occurs again, here due to soliton formation, because self-phase modulation comes into balance with anomalous dispersion[1,38]. At that point, typically also soliton fission and Raman red-shifting occurs, and dispersive waves are generated. However, fission does not increase the bandwidth[1,36], dispersive waves generate only disconnected components of narrowband radiation, and Raman conversion red-shifts the frequency bandwidth, without increasing it.

It was shown that the bandwidth of dispersive waves can be broadened in the anomalous dispersion regime by waveguide tapering[28,39,40]. However, unavoidable and intrinsic bandwidth stagnation occurs by the taper geometry itself when the anomalous dispersion is no longer present and by the early onset of increasing pulse duration and decreasing peak power, introduced by these schemes[39,40] after an initial nonlinear pulse compression.

For avoiding both pulse stretching and soliton formation, it seems desirable to provide zero group velocity dispersion (GVD), to keep self-phase modulation ongoing and to increase the bandwidth proportional to the interaction length. However, this would not yield the maximum possible broadening with length, because newly generated spectral bandwidth is not used for pulse compression during propagation.

We present a novel concept (see Fig.1) based on *repeatedly alternating the sign of group velocity dispersion* along propagation in a segmented medium. The alternation overcomes spectral stagnation, by both countering temporal broadening and disrupting soliton formation, and thereby lowers the input peak power and pulse energy required for supercontinuum generation. In its simplest setting, self-phase modulation in normal dispersive segments generates spectral broadening and subsequent segments with anomalous dispersion make use of the newly generated bandwidth for pulse re-compression and peak power increase, which lets spectral broadening continue. An approximate analysis reveals close-to-exponential growth of bandwidth vs propagation at weak dispersion and linear growth at strong dispersion. This scaling with interaction length enables to generate a given spectral bandwidth with substantially reduced pulse energy and peak power as compared to conventional supercontinuum generation based on uniform dispersion.

## Results

**Basic dynamics with alternating dispersion**
Figure 1 shows calculated developments of the optical spectrum and pulse along propagation for a dispersion alternated waveguide (Fig. 1A) made of normal dispersive (ND) segments (positive group velocity dispersion) and anomalous dispersive (AD) segments (negative group velocity dispersion). For unveiling the novel spectral dynamics caused by repeatedly alternating dispersion, the calculations retain only the two main physical processes, which are self-phase modulation (SPM) causing coherent spectral broadening and second-order dispersion responsible for temporal stretching or compressing the pulse. For convenience, we take transform limited Gaussian pulses as input and consider spectral broadening only in the ND segments. Other situations, e.g., non-Gaussian input pulses, spectral broadening in both types of segments, higher-order dispersion, or propagation loss are taken into account as well, when comparing with experimental data as in Figs. 2 and 3.

Figure 1 B shows the increase in spectral bandwidth vs propagation coordinate as obtained by numerical integration of the nonlinear Schrödinger equation (NLSE) using parameters as in telecom ND and AD fiber (Supplementary Section 1). It can be seen that there is an initial spectral broadening through SPM in the first ND segment, but broadening stagnates beyond approximately $z_1$=15 cm. The reason for stagnation in this ND segment is loss of peak intensity and increased duration by normal dispersive temporal broadening of the pulse (Fig. 1C).

To re-initiate spectral broadening by re-establishing high peak intensity, the pulse is temporally re-compressed in an AD segment back to its transform limit (between $z_1$ and $z_2$=30 cm).

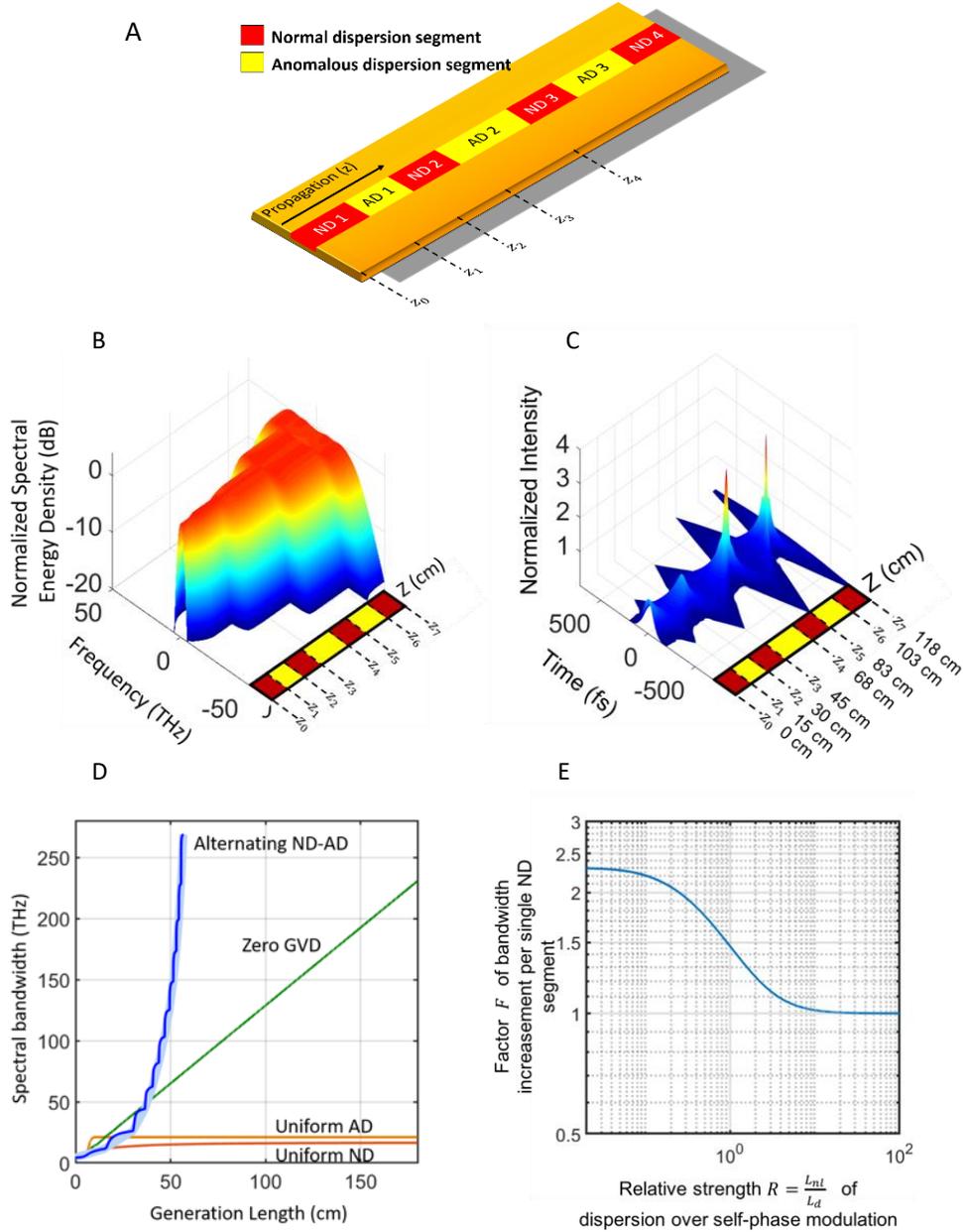

**Figure 1:** Supercontinuum generation (SCG) using alternating dispersion **A**: Waveguide with segments of normal dispersion (ND, red) and anomalous dispersion (AD, yellow). The basic pulse and spectral propagation dynamics is modelled with Gaussian input pulses using parameters as in telecom fiber (Supplementary Section
3

1), accounting for second-order dispersion in ND and AD segments, and self-phase modulation in the ND segments. **B:** Normalized energy density spectrum vs. propagation with alternating dispersion. **C:** Intensity and shape of the pulse vs propagation. Undesired pulse stretching in the bandwidth generating (ND) segments is repeatedly compensated by anomalous dispersive (AD) segments. **D:** Spectral bandwidth of supercontinuum generation in alternating dispersion (ND-AD) compared to zero uniform dispersion (zero GVD), and to uniform normal and anomalous dispersion (uniform AD, uniform ND). For direct comparison with the uniform waveguides the graph omits non-generating AD segments. With the chosen fiber parameters and input pulse, alternating dispersion yields close-to-exponential growth (blue-shaded exponential fit curve under blue trace). Supercontinuum generation in uniform ND and AD exhibits spectral stagnation. **E:** Calculated spectral bandwidth increasement factor, $F$, in single ND segments vs the relative strength of dispersion and self-phase modulation, $R$, in terms of the ratio of nonlinear length to dispersive length ($R = L_{nl}/L_d$).

The pulse at $z_2$ is shorter and has a higher intensity than the incident pulse ($z_0 = 0$ cm) because SPM in the first ND segment has increased the bandwidth above that of the input pulse. The consequence of higher intensity and shorter pulse duration is that SPM in the second ND segment yields more additional bandwidth (between $z_2$ and $z_3$) than what was gained in the first ND segment (between $z_0$ and $z_1$). Also, the rate of bandwidth growth and dispersion are higher, so that spectral stagnation is reached after a shorter propagation distance than in the first ND segment. Following this concept, further dispersion alternating segments keep the bandwidth growing and let the pulse become shorter as shown.

Fig. 1 D provides a bandwidth comparison with conventional supercontinuum generation (uniform AD, uniform ND) where the growth of bandwidth stagnates. In alternating dispersion (ND-AD), bandwidth generation remains ongoing with each next ND segment (AD segments omitted in for direct comparison with generation in uniform AD and ND media). With the chosen fiber and pulse parameters, the spectral bandwidth grows close-to-exponential (see blue-shaded exponential fit curve) and reaches about 200 THz via SPM in 50 cm of bandwidth generating ND segments. In uniformly dispersive fiber (ND or AD) there is spectral stagnation, limiting the bandwidth to 15 THz (after 20 cm) and 20 THz (after 10 cm), respectively.

The bandwidth at stagnation in uniform dispersion may be increased as well, however, this requires significantly increased power, because the bandwidth grows approximately only with the square-root of the input power[1,4,36]. For instance, generating 200 THz in uniform dispersion with the fiber parameters used in Fig.1, would require about 180-times higher power (with ND), and 100-times higher power with uniform AD. The bandwidth comparison in Fig. 1 D is thus indicating that alternating dispersion allows to significantly reduce the power required for generating a given bandwidth.

**Scaling of bandwidth with segment number and input power**

Many applications benefit from a broad spectral bandwidth particularly if there is significant power also at the edges of the specified bandwidth, i.e., if spectra are approximately flat in the range of interest[36,37]. The standard definition of bandwidth for supercontinuum generation leads to wider bandwidth values as it accepts low power at the edges, typically at the -30-dB level. To follow a conservative definition of bandwidth addressing approximately flat spectra, we chose as bandwidth, $\Delta v$, the full width frequency bandwidth at the 1/e-level of the energy spectral density.

To provide spectra with minimum interference structure[4,36,37], we continue considering that spectral broadening is generated in the ND segments with Gaussian pulses, and that the AD segments just re-compress the pulse to the transform limited duration, $\Delta t \cdot \Delta v = 2/\pi$, where $\Delta t$ is the 1/e half-width duration.

For calculating the bandwidth increase in a particular ND segment when a transform limited pulse enters the segment, it is important to realize that the achievable broadening is ruled by only two parameters. The first is the nonlinear length, $L_{nl} = 1/(\gamma P)$, depending reciprocally on the peak power $P$, where $\gamma$ is the nonlinear coefficient of the ND segment. $L_{nl}$ denotes the propagation length where the bandwidth increases by a factor $\sqrt{5}$ if dispersion were absent[41]. The second parameter is the dispersion length, $L_d = \Delta t^2/|\beta_2|$ depending quadratically on the pulse duration, where $\beta_2$ is the coefficient for second-order (group velocity) dispersion[41]. $L_d$ denotes the propagation length after which the pulse duration increases by a factor of $\sqrt{2}$ if SPM were absent.

Given a certain length of the ND segment, it is the ratio of $L_{nl}$ and $L_d$ which determines the bandwidth gained in that segment. For instance, if dispersion is weak compared to spectral broadening, i.e., if the ratio $R = L_d/L_{nl}$ is small ($R \ll 1$), one expects the bandwidth to increase by a factor of $F = \sqrt{5}$ if the segment length is chosen equal to the nonlinear length. If dispersion is strong vs SPM, due to a short pulse duration and thus a wide spectrum ($R \gg 1$), there will still be broadening in the beginning of the segment, but the bandwidth increasement factor will only be slightly above unity, $F \gtrsim 1$.

To quantify the relation between the ratio $R$ and spectral broadening in an ND segment, we devise a lower-bound calculation of the bandwidth increasement factor, $F(R)$, and display the result in Fig.1E (see derivation of $F(R)$ in Supplementary Section 2). For the calculation, we chose the length of the ND segment to be its nonlinear length, to encompass all nonlinear generation up to the point where the bandwidth stops increasing. It can be seen that $F$ varies with $R$ as expected, i.e., having its values close to $\sqrt{5}$ at small $R$ and reducing asymptotically to unity towards increasing $R$.

We note that $F(R)$ does not only determine the broadening in a single ND segment, but it determines also the broadening in all following ND segments. To illustrate this, we recall Fig.1 where, due to recompression in AD segments, the peak power increases and pulse duration shortens before entering a next ND segment. This causes $R$ to increase with each ND segment since $R \propto 1/\Delta t$. More specifically, because the $R$-value of a previous ND segment determines the bandwidth increasement factor $F$ in the according segment (see Fig. 1 E), it also determines the bandwidth, transform limited pulse duration and peak power in front of the next ND segment. The peak power and pulse duration then set the $R$-value of the next ND segment and, via $F(R)$, the next bandwidth increase. Labelling the ND segment number with $p$, the recursive dependence of $R_p$ on $R_{p-1}$ is why the first segment, $R_1$, pre-determines the bandwidth growth along the entire structure.

The bandwidth at the end of the structure with a number of $n$ ND segments, if $\Delta \nu_0$ is the frequency bandwidth of the incident pulse, can then be expressed as a product of broadening factors,

$$\Delta \nu_n = \Delta \nu_0 \cdot \prod_{p=1}^n F_p \qquad . \qquad \text{Eq.1}$$

Here $F_p \equiv \Delta \nu_p / \Delta \nu_{p-1}$ is the bandwidth increasement factor of the $p^{th}$ ND segment, where $\Delta \nu_{p-1}$ is the bandwidth in front of the segment and $\Delta \nu_p$ is the bandwidth behind the segment.

Particularly strong growth is obtained from Eq.1 if generation remains in the range of relatively weak dispersion ($R \leq 1$)). In this case $F_P$ remains close to its maximum value ($F_P(0) = \sqrt{5} \approx 2.23$) or decreases only slowly with segment number. The bandwidth growth then remains approximately exponential with the number of ND segments, resulting in

$$\Delta \nu_n = \Delta \nu_0 \cdot F^n \qquad , \qquad \text{Eq. 2}$$

as is found also numerically for the example in Fig. 1D. In the other limit of strong dispersion ($R \gg 1$), we find (more details in Supplementary Section 3) that the decrease in $F_P$ with each additional ND segment reduces Eq.1 to a linear growth of bandwidth with the number of ND segments,

$$\Delta\nu_n \approx n\left[g_o\gamma_2\frac{E}{4|\beta_2|}\right] + \Delta\nu_0 \qquad , \qquad \text{Eq. 3}$$

with $g_o \approx 0.81$ being a constant related to the Gaussian pulse shape, and with $E = \sqrt{\pi}\, P \cdot \Delta t$ being the pulse energy.

Equations 2 and 3 display that the basic dynamics of supercontinuum generation in alternating dispersion is a growth of spectral bandwidth vs propagation length, while with uniform dispersion spectral stagnation occurs. The growth may remain close to exponential if $R$ remains small throughout all segments, show transition from exponential to linear growth at around a certain segment number $n_T$ where $R \approx 1$, or is close to linear if dispersion is strong in the first and thus also in all segments ($R_1 \gg 1$).

Using these properties, we find approximate bandwidth-to-input peak power scaling laws (more details in Supplementary Section 3) shown in Table 1. In uniform dispersion there is only a square-root growth of bandwidth vs input peak power. With alternating dispersion, the bandwidth increases in a mixture of exponential and linear growth vs input peak power, where the slope and exponent can be increased monotonically with the total number of ND segments. Correspondingly, the generated bandwidth changes much more strongly with input peak power in the alternating dispersion structure. Therefore, the power required to achieve a given bandwidth can be much reduced by increasing the number of ND segments.

| Relative strength of dispersion over self-phase modulation in first segment | Scaling of bandwidth with input peak power using alternating dispersion | Scaling of bandwidth with uniform dispersion |
|---|---|---|
| $R_1 > 1$ to $R_1 \lesssim 1$ | $\Delta\nu_n \propto (c_T\sqrt{P_O})^{n_t-1} + (n - n_t + 1)P_O$ | $\Delta\nu_n \propto P_o^{\frac{1}{2}}$ |
| $R_1 \ll 1$ | $\Delta\nu_n \propto [c_T{}^{n_t-1} + (n - n_t + 1)]P_O$ | $\Delta\nu_n \propto P_o^{\frac{1}{2}}$ |

**Table 1:** Scaling of bandwidth vs input peak power, $P_0$, and number of segments, $n$, compared to power scaling in uniform dispersion. Alternating dispersion yields a mixture of linear and exponential growth vs the input peak power, depending on the relative strength of dispersion over self-phase modulation in the first ND section (expressed as the ratio of nonlinear length to dispersive length, $R^{(1)} = L_{nl}^{(1)}/L_d^{(1)}$. $n_T$ is the segment number where exponential growth transits into linear growth (where $R \approx 1$), and $c_T$ is a constant (see Supplementary Section 3 regarding the definition of $n_T$ and $c_T$)

Summarizing the analytical results, repeatedly sign-alternating dispersion offers new spectral bandwidth dynamics that are fundamentally different than conventional supercontinuum generation schemes in uniform dispersion or tapered waveguides. These nonconventional dynamics increase the ratio of bandwidth to input peak power while the dynamics follows different regimes across the propagation coordinate or when changing the input peak power.

**Experimental results**
For demonstrating increased spectral broadening with repeated sign-alternating dispersion, we use supercontinuum generation with ultra-short optical pulses injected into a repeatedly dispersion sign-alternated optical fiber structure shown in Fig. 2A. The incident pulses have 74 fs FWHM pulse duration at 50 mW average power and 79.9 MHz repetition rate (central wavelength of 1560 nm). The pulses are provided by a mode-locked Erbium doped fiber laser, amplifier and a temporal compressor system. The power coupled into the fiber is 35.6 mW (446 pJ pulse energy) and is held constant throughout the main experiments.

The dispersion alternated fiber structure is selected to comprise alternating segments of standard single-mode ND and AD telecom fiber with well-characterized linear and nonlinear properties (see Fig. 2B and parameters in Supplementary Section 4) because this enables numerical modeling with high reliability. We note that the wavelength range across which the sign of the dispersion can be alternated and the concept can be applied, is ultimately given by the ND and AD zero-dispersion wavelengths indicated with dashed lines in Fig. 2B.

This proof of concept experiment was designed to demonstrate that our concept works in real cases where there is uncompensated higher order dispersion in the segments, limited sign-inverted bandwidth, losses and low nonlinearity.

In the experiment, ND and AD segments are added sequentially, with the choice of segment length guided by intermittent recordings of the energy spectrum and the pulse duration (see Methods). Due to the almost five-times higher nonlinear coefficient in the ND fiber compared to the AD fiber, spectral broadening comes almost entirely from the ND segments, similar to the setting used for illustrating the basic dynamics in Fig.1.

Figure 2C shows supercontinuum spectra measured behind each added ND segment. Starting with the injected light, the spectral bandwidth increases with each ND segment. For better analysis, Fig. 2D displays the measured bandwidth growth vs segment number (red dots). To identify the regime of growth, we determine the experimental ratio $R = L_{nl}/L_d$ using the measured peak power and pulse duration. We find $R \leq 1$ for the first two segments, suggesting that the initial growth would be closer to exponential growth, while the third and fourth segments would introduce a transition towards linear growth. This is confirmed by finding a slightly upwards curved growth along the first three data points, followed by a transit towards more linear growth with the remaining segments.

The maximum number of ND segments before no further spectral broadening occurs is seen to be four. This can be attributed to the finite width of the spectral interval across which dispersion parameters show opposite signs (Fig. 2B), i.e., the range in which alternating dispersion can be applied with the used fibers.

Figure 2E shows supercontinuum spectra generated with dispersion alternated fiber with four ND segment (trace 1). For direct bandwidth comparison, we generate supercontinuum also in uniform fiber (trace 2 and 3 for uniform ND and AD fiber, respectively), using the same input pulse parameters and fiber length. The comparison shows that alternating dispersion increases the full 1/e bandwidth by a factor of 1.4 and 2.5 with regard to uniform ND and AD fiber, respectively. The full-width at -30 dB increases by a factor of 2.5 and 3.1, respectively. The spectral profile associated with the dispersion alternated fiber is free of large spectral modulations, which is due to spectral generation occurring predominantly in the ND segments. The dashed traces are numerical solutions of the generalized nonlinear Schrödinger equation, accounting for all orders of dispersion, the experimental pulse shape, loss between segments, and also weak SPM in the AD segments (see parameters in Supplementary Section 4). It can be seen that there is very good agreement with the experimental data in all cases, i.e., for supercontinuum generation in uniform ND and AD fiber, and also for the dispersion alternated fiber.

Figures 2F compares the development of the pulse duration with according measurements of the intensity autocorrelation (Fig. 2G). We find good agreement between the modelled oscillation of the pulse duration with that of the autocorrelation measurements.

Motivated by the agreement and given the spectral bandwidth generated with dispersion alternated fiber, we use the model to extrapolate the pulse energy needed to generate the same bandwidth with uniform ND fiber. The model predicts that approximately one order of magnitude higher pulse energy is required with uniform dispersion. We note that this factor is a conservative number due to splice loss between segments (about 8% per splice) which lowers the intensity available for SPM towards the later segments. Lower required powers are expected with post-processing the splices. However, the presence of high splice losses show that our concept is robust to this influence. The compression to ultrashort pulse durations show that the spectral phase coherence is maintained in the alternated fiber. Closer investigation of Shot-to-shot coherence will be the subject of future work. The proper working of the scheme also in the presence of substantial non-compensated third and higher-order dispersion (see Fig. 2B) indicates a certain robustness of sign-alternating dispersion vs higher-order dispersion, as we estimate a maximum deviation from a parabolic spectral phase profile by at most 13% within the 1/e bandwidth. We obtain this spectral phase profile estimate through the autocorrelated pulse duration and associated bandwidth in front of the last AD segment and directly behind it. A near parabolic spectral phase profile is usually characteristic of spectral generation in the normal dispersion regime from the interplay of SPM and dispersion[36].

The experiment verifies the theory of repeated sign-alternating dispersion, since there is a large bandwidth enhancement relative to uniform dispersion cases. This in turn indicates that dispersion alternating provides a considerably larger bandwidth to input peak power ratio. Additionally, the dispersion alternating fiber keeps advantages of normal dispersion supercontinuum generation[4,36,37] seen by a spectrum free of large modulations and the high temporal compressibility of the pulses.

While the experiment verifies the functioning of our concept, it is a proof-of-concept experiment and further improvements (minimize splice losses, increase nonlinearities, etc.) will promise to further increase the bandwidth generation. However, the experiment provides the insight that repeatedly sign-alternating dispersion can remove the need to use maximally nonlinear waveguides, since the alternation substantially increases spectral broadening in a waveguide system with a low nonlinear coefficient relative to nonlinear fiber (e.g., PCFs). A wider range of waveguide platforms and materials is thus made available for supercontinuum generation.

The experiment also shows that our concept removes the need to use ultrashort or high peak power pulses, thus enabling supercontinuum generation with a wider range of lasers. This should enable, for example, implementation of supercontinuum generation with chip integrated lasers that have lower peak power, e.g., due to a high repetition rate.

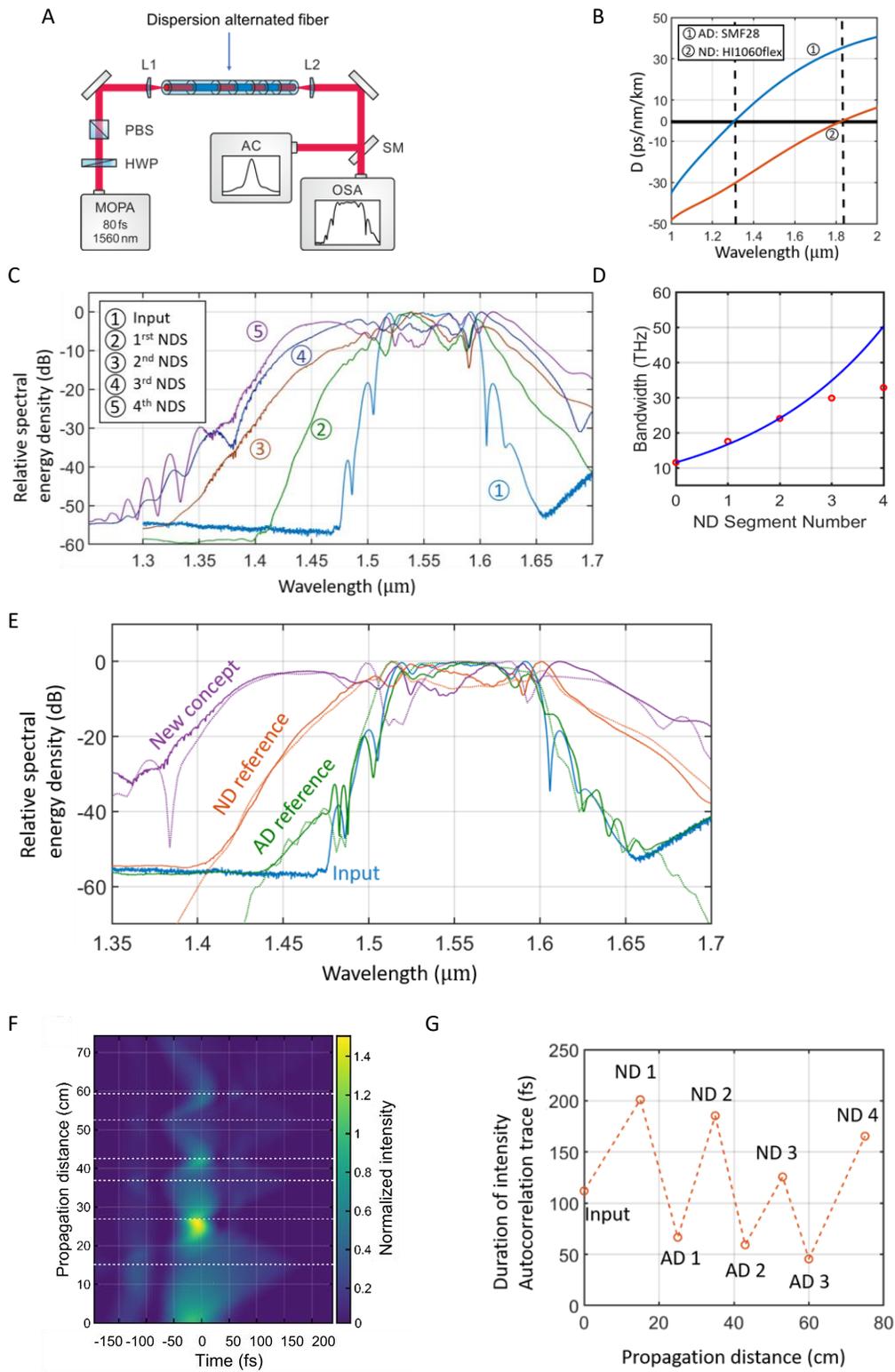

**Figure 2: A**) Experimental setup for supercontinuum generation (SCG) in optical fiber with alternating normal dispersion (ND) and anomalous dispersion (AD). HWP: half-wave plate. PBS: polarization beam splitter. L1, L2: lenses. SM: switchable mirror. OSA: optical spectrum analyzer. AC: autocorrelator. **B**) Dispersion parameter, $D = \beta_2 \cdot (2\pi c/\lambda^2)$, vs wavelength of the AD and ND fiber segments. The wavelengths of zero second-order dispersion are indicated with dotted vertical lines. **C**) Relative spectral energy density of supercontinuum generation measured behind each ND segment (NDS) in the alternating waveguide structure. The spectral bandwidth grows in steps with added ND segments. **D**) Measured bandwidth vs ND segment number retrieved from the spectra in C. The measured pulse duration and peak power entering the segments, used to determine the $R$-value for each segment (not separately shown), suggest an approximately exponential growth via the first two segments (compare to exponential curve beginning at input bandwidth). Thereafter, weaker growth transiting to linear is expected. **E**) Measured normalized spectral energy density at the end of the dispersion alternating fiber (violet trace). For direct bandwidth comparison, measured supercontinuum spectra obtained with uniform ND fiber (brown trace) and AD fiber (green) of the same length are shown, generated with the same input pulse parameters. The dotted traces show theoretical spectra obtained with the generalized nonlinear Schrödinger equation, using the experimental input spectrum, containing the full fiber dispersion (B) and weak SPM also in the AD segments. **F**) Calculated normalized intensity (color coded) vs time (horizontal axis) and vs the propagation coordinate (vertical axis). Dotted white lines indicate the transitions between ND and AD segments. **G**) Measured width of auto-correlation trace (FWHM) versus propagation distance in the dispersion alternated fiber. The dotted trace connects experimental data. The re-compressed pulse is becoming shorter after each further AD segment.

**Integrated optical waveguides with alternating dispersion**

Particularly promising is supercontinuum generation in integrated optical waveguides because tight guiding reduces the required input power to provide high intensities. Further advantages are that lithographic fabrication methods provide increased freedom and precision in dispersion engineering[42]. For instance, different from fiber splicing, transitions between segments can be shaped as desired, and also the spectral shape of ND and AD dispersion may be designed to widen the wavelength range between ND and AD zero dispersion wavelengths.

To investigate the options for additional spectral broadening and reduced peak power in such systems, we use the generalized nonlinear Schrödinger equation[11,41,43] to model supercontinuum generation in silicon nitride integrated optical waveguides. Figure 3A compares generation in uniform anomalous dispersion with that in sign-alternating dispersion (see parameters in Supplementary Section 5). With uniform dispersion (see Fig. 3B, red curve), the assumed input pulse energy and duration (450 pJ and 120 fs), the shape of the calculated spectrum (Fig. 3A, blue trace) and the 1/e full bandwidth (21 mm) are comparable with previous modelling that matched according experiments[10].

For modelling with alternating dispersion, we select four ND segments (see dispersion in Fig 3B, blue curve) and three AD segments, beginning with an ND segment (parameters in Supplementary Section 5). We note that in this structure, different than with the investigated fibers, both types of segments contribute to spectral generation and high-order dispersion is included.

The second trace in Fig. 3A (red) shows the spectrum obtained with the alternating structure and 80 pJ input pulse energy. Both approaches yield approximately the same -30-dB bandwidth, however, the input pulse energy is about 5.6-times lower with alternating dispersion. A second, difference is the much larger 1/e-bandwidth (550nm), which is a factor of about 26 wider. While the generated bandwidth is both larger and flatter in the alternating waveguide than in the uniform dispersion case, effects such as modulation instability, coherence degradation and complex spectral phase profiles may still be present due to a notable fraction of spectral generation in the anomalous dispersion segments. The complex dynamics introduced here will be subject of future work.

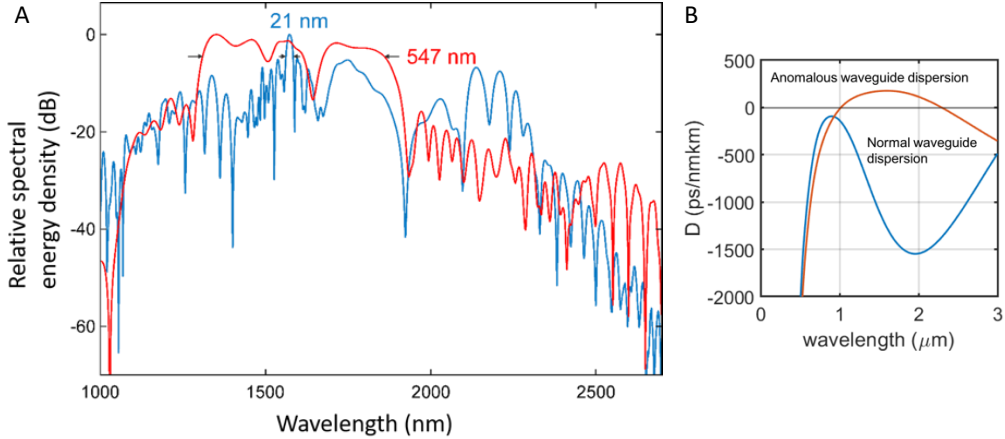

**Figure 3: A)** Normalized spectral energy density (dB) versus wavelength compared for the same length of uniform normal dispersive waveguide (blue trace), and the spectrum obtained with alternating dispersion (red trace). **B)** Dispersion parameter curves for the AD (red) and ND (blue) segments used in the integrated optical waveguide (parameters see Supplementary Section 5).

## Conclusions

Repeatedly alternating the sign of dispersion in supercontinuum generation is a novel and highly promising concept for increasing the generated spectral bandwidth and for reducing the required input peak power or pulse energy. Even in a situation where the -30-dB bandwidth is already extremely wide, the flat part of the spectrum specified via the 1/e bandwidth can be much increased and, simultaneously, the required power can be much reduced.

We note that, in the limit of linear optics, dispersive sign alternation is well known in optical fiber communications for preventing pulse stretching, and to ensure that the dynamics remains in the linear regime, to avoid cross-talk of pulses[44,45]. Periodic dispersion oscillation is also known for providing quasi-phase matching (QPM) of dispersive waves with solitons, for enhancing spectrally narrowband modulation instability, or to generate soliton trains[46,47].

The concept of alternating dispersion presented here goes far beyond, because it offers wider bandwidth at reduced power. Due to its generality, we expect that the concept can have important impact in a large variety of waveguiding systems that aim at coherent generation of broadband optical light fields. Examples may include photonic crystal fibers[48], low-power broadband generation for dual-comb applications in the mid-infrared[49], on-chip generation of frequency combs[50], and coherent spectral broadening in liquid[51] and gaseous media[52].

The basic concept as presented, encourages further investigations into various directions. For instance, spectrally shaping normal and anomalous dispersion via dispersion engineering in photonics waveguides can be explored to extend the range between zero-dispersion wavelengths. Emphasizing self-phase modulation in normal dispersive segments, rather than in anomalous dispersion, might be used to reduce undesired spectral structure, nonlinear phase spectra, and noise from modulation instability, as is found in supercontinuum generation with anomalous dispersion[4,36,37].

For optimum results, i.e., maximally wide spectra at lowest possible power, alternating dispersion requires an optimum choice of the segment lengths. Sets of adjacent waveguide structures on chip may be used to cope with wider variations of the input power or pulse shape. Nevertheless, because alternating dispersion always counteracts the detrimental effects of uniform dispersion, we observe that noticeably increased bandwidth and lowered power requirements can be obtained even with non-optimally matched segment lengths, for instance with simple periodic sign-alternation.

## Methods

To assemble the dispersion alternated fiber, we use segments of standard single-mode doped-silica step-index optical fiber (Corning Hi1060flex for ND, Corning SMF28 for AD). The structure is assembled using fiber splicing and cut-back, starting with a 25 cm piece of ND fiber. While measuring the power spectrum at the fiber output with an optical spectrum analyzer, the length is cut back until a slight reduction of the spectral width becomes noticeable at the $-30$ dB level. Terminating the first segment at this length ensures that the spectral generation is ongoing throughout the entire segment but stagnates at the end of the segment. The cutback also removes further pulse defocusing by second and higher-order dispersion while no further broadening takes place. Removing higher-order phase contributions improves pulse compression in the following AD segment because the AD fiber compensates for second-order dispersion but, in our demonstration, cannot compensate for the high-order dispersion of the ND fiber.

To assemble the next segment, a 20 cm piece of AD fiber is spliced, the splice loss is measured with a power meter and the pulse duration is measured with an intensity auto-correlator (APE, Pulse Check). The AD fiber segment is cut down until the autocorrelation trace (AC) indicates a minimum pulse duration, i.e., that the pulse is closest to its transform limit at the end of the AD segment. The procedure is then repeated with the next pieces of ND and AD fibers. The lengths of the fiber segments obtained with this method and the measured splice loss are summarized in Supplementary Section 4.

## Acknowledgements

The authors thank D. Marpaung and F. Schepers for helpful discussions in writing the manuscript. The authors thank Toptica GmBH for lending the pump laser system. C.F. and K.-J. B. acknowledge support via the International Strategic Partnership Program of the University of Twente for the collaboration between both involved universities (UT and WWU). K.-J. B. and H. Z. thank funding from the MESA+ Institute of Nanotechnology within the grant "Ultrafast switching of higher-dimensional information in silicon nanostructures".

**Contributions**

H.Z. conceived and developed the concept, the theory and numerical GNLSE model and designed the experimental methodology. N.L., C.F. designed and built the experimental setup with help from H.Z.. N.L., H.Z. conducted the experiment. H.Z., N.L. analyzed the experimental data. N.L. calculated the fiber and waveguide dispersion. K.-J. B., H.Z., C.F., N.L. wrote the manuscript. H.Z. wrote the supplementary section. K.-J. B. and C.F. supervised the work. All authors contributed to the scientific discussion.

**Competing Interests**

H.Z and K.-J. B. are co-inventors of a pending filed patent owned by University of Twente of which a part covers the main idea of this work. The patent application no. is 16/204,614.

# Supplementary Information

Haider Zia[1*], Niklas M. Lüpken[2], Tim Hellwig [2], Carsten Fallnich[2,1], Klaus-J. Boller[1,2]

[1] University of Twente, Department Science & Technology, Laser Physics and Nonlinear Optics Group, MESA+ Research Institute for Nanotechnology, Enschede 7500 AE, The Netherlands

[2] University of Münster, Institute of Applied Physics, Corrensstraße 2, 48149 Münster, Germany

*h.zia@utwente.nl

## Supplementary Section 1

In the following table S1, S2 the parameters are shown that have been used for the simulations presented in Fig. 1B and C and Fig. 1D.

Table S1. Parameters for Simulations of Fig. 1

| Parameter | Fig. 1B, 1C |
|---|---|
| **Input pulse energy** | 1 nJ |
| **Input power $e^{-1}$ half duration** | 72 fs |
| **ND $\gamma$** | $5.2E-3\ W^{-1}m$ |
| **AD $\gamma$** | 0 |
| **ND $\beta_2$** | $2.61E4\ fs^2 m^{-1}$ |
| **ND segment length** | 15 cm |
| **Total ND length** | 90 cm |
| **AD $\beta_2$** | $-1.89E4\ fs^2 m^{-1}$ |
| **AD segment lengths** | 15.01, 23.30, 19.45 cm |

Table S2. Parameters for Simulations of Fig. 1D

| Parameter | ND | AD | Alternating sructure | 0 GVD |
|---|---|---|---|---|
| **Input pulse energy** | 1 nJ | 1 nJ | 1 nJ | 1 nJ |
| **Input intensity $e^{-1}$ half duration** | 72 fs | 72 fs | 72 fs | 72 fs |
| **ND $\gamma$** | $5.2E-3\ W^{-1}m$ | Not used | $5.2E-3\ W^{-1}m$ | $5.2E-3\ W^{-1}m$ |

| AD $\gamma$ | Not used | $5.2E-3\ W^{-1}m$ | 0 | Not used |
| --- | --- | --- | --- | --- |
| ND $\beta_2$ | $2.61E4\ fs^2m^{-1}$ | Not used | $2.61E4\ fs^2m^{-1}$ | 0 |
| ND segment length | Not used | Not used | Taken at 90% spectral saturation | Not used |
| Total ND length | 180 cm | 180 cm | 56 cm | 180 cm |
| AD $\beta_2$ | Not used | $-2.61E4\ fs^2m^{-1}$ | Pulses were compressed to transform limit at entrance of ND segments | Not used |
| AD segment lengths | Not used | Not used | Not used | Not used |
| Total AD length | Not used | 180 cm | Not used | Not used |

## Supplementary Section 2

**Bandwidth Increasement Factor**

In this supplementary section we present a derivation of the analytic expressions used to calculate the lower bound increasement factor, $F$, presented in Fig. 1E for SPM in normal dispersion (ND). In general, it is hard to avoid full numerical simulations for systems based even on the simplest generalized nonlinear Schroedinger equations (GNLSE) [1] and input conditions, such as with Gaussian pulses. The following analytic expressions aim on providing more insight, give a lower bound estimate, clarify how the spectral bandwidth increases across the sign-alternating dispersion structures and they allow for the design of such structures to meet a certain bandwidth criterion, without resorting to proofs by extensive numerical simulations based on parameter variation. In the main text, we use the frequency bandwidth $\Delta \nu$ for easy comparison with experiments. In the following derivations we use, for convenience, the angular frequency, $\omega = 2\pi \cdot \nu$ and $\Delta\omega = 2\pi \cdot \Delta\nu$. The bandwidth increasement factor, $F$, remains independent of this choice.

We assume, as input, transform limited Gaussian pulses, although our expressions allow for chirped inputs as well. Our lower bound estimate treats dispersion and SPM up to second order, so the pulse remains Gaussian as it propagates. It can be shown that this is sufficient for tracing the lower bound 1/e bandwidth evolution for Gaussian inputs. It can also be shown, that even with higher order dispersion, taking the largest absolute group velocity dispersion (GVD) value within the considered range between two zero-dispersion frequencies ($\Delta\omega_r$) for our method, would guarantee a lower bound estimate. Our lower bound estimate lies closest to actual values for systems where the GVD is a slow varying function, i.e., $\beta_2 \Delta\omega_r^2 > \beta_3 \Delta\omega_r^3$, with $\beta_2, \beta_3$ being the second and third order dispersion coefficients at the carrier frequency.

Under constant second order dispersion, propagation along z leads to a spectrum with quadratic phase [1]:

$$E(\omega, z) = A_o e^{-[\Delta t_0^2]\frac{\omega^2}{2}} e^{i\frac{\beta_2}{2}(\omega)^2 z} \quad , \qquad\qquad \text{S1}$$

where, $A_o e^{-[\Delta t_0^2]\frac{(\omega)^2}{2}}$ is the initial transform limited input with an $e^{-1}$ intensity half duration of $\Delta t_0$. $A_o$ is a constant related to the peak input amplitude $\omega = 2\pi(\nu - \nu_0)$, with $\nu_o$ being the central frequency.

In the time domain, Eq. S1 represents a temporally chirped Gaussian pulse with an $e^{-1}$ intensity half duration $\Delta t$ that grows along $z$:

$$\Delta t(z) = \left[1 + \left(\frac{z}{L_D}\right)^2\right]^{\frac{1}{2}} \Delta t_0 \quad \text{with } L_d = \frac{\Delta t_0^2}{|\beta_2|}\ . \qquad\qquad \text{S2}$$

Describing SPM spectral generation in the time domain, neglecting GVD, yields that the temporal Gaussian function is multiplied by a nonlinear (i.e., power dependent) temporal phase function represented as:

$$E(t,z) = \sqrt{P_o}\, e^{-\frac{t^2}{2\Delta t_0^{\,2}}} e^{i\gamma P z} \qquad , \qquad \text{S3}$$

where $P_o$ is the peak power of the pulse. $P$ is the time dependent power of the input pulse given as $P = P_o e^{-\frac{t^2}{\Delta t_0^{\,2}}}$. We use a Taylor series expansion of $P$ up to the second order term, yielding:

$$P = P_o \left[1 - \frac{t^2}{\Delta t_0^{\,2}}\right] \qquad . \qquad \text{S4}$$

Substituting this into S3 yields:

$$E(t,z) \approx \sqrt{P_o}\, e^{-\frac{t^2}{2\Delta t_0^{\,2}}} e^{i\gamma P_o \left[1 - \frac{t^2}{\Delta t_0^{\,2}}\right] z} \qquad . \qquad \text{S5}$$

The first term in the square bracket can be omitted without loss of generality. It can be seen that the expression becomes formally analogous to Eq. S1 with $t$ swapped for $\omega$. Therefore, using the analogous derivation, as for Eq. S2, $\Delta\omega(z)$, the $e^{-1}$ spectral energy density value half angular bandwidth is given as:

$$\Delta\omega(z) = \left[1 + \left(\frac{z}{L_N}\right)^2\right]^{\frac{1}{2}} \Delta\omega_0 \qquad . \qquad \text{S6}$$

To derive how $L_N$ depends on $P_0$, we find the analogous quantities to the expression of $L_d$. The $\Delta t_0^{\,2}$ in the expression for $L_d$ is then replaced by $\Delta t_0^{\,-2}$ and the $|\beta_2|$ term is replaced by $2\left[\frac{\gamma P_o}{\Delta t_0^{\,2}}\right]$ for the corresponding expression for $L_N$. Then the expression for $L_N$ is $L_N = \frac{\Delta t_0^{\,-2}}{2\left[\frac{\gamma P_o}{\Delta t_0^{\,2}}\right]} = \frac{1}{2\gamma P_o}$. In analogy to $L_d$, $L_N$ represents the length necessary for the spectral bandwidth to increase by a factor of $\sqrt{2}$ (without the inclusion of dispersive effects). Note that $L_N$ is not the nonlinear length $L_{nl}$ as used in the main text, but related to it as $L_N = L_{nl}/2$. We define $L_N$ as such to reduce the complexity of Eq. S6 and expressions that follow.

The final, full expression for the bandwidth increasement factor, $F$, obtained from Eqs. S2 and S6 as shown in the proof section below, is:

$$F = \left[1 + \left[\frac{\Delta\tau_{max}^{R2}}{\Delta\tau_{max}^{R1}}\right]^2 \left(\left[\frac{\Delta\tau_{max}^{R1}}{\Delta t_0}\right]^2 + \left[\frac{\Delta t_0}{\Delta\tau_{max}^{R1}}\right]^2 - \left[\frac{\Delta\tau_{max}^{R1}}{\Delta\tau_{max}^{R2}}\right]^2 + 2\frac{\Delta t_0}{\Delta\tau_{max}^{R1}}\left[\left(\frac{\Delta\tau_{max}^{R1}}{\Delta t_0}\right)^2 - 1\right]^{\frac{1}{2}}\right) + \right.$$

$$\left. 2\left[\frac{\Delta t_0}{\Delta\tau_{max}^{R1}}\right]\left(\left[\frac{\Delta\tau_{max}^{R1}}{\Delta t_0}\right]^2 + \left[\frac{\Delta t_0}{\Delta\tau_{max}^{R1}}\right]^2 - \left[\frac{\Delta\tau_{max}^{R1}}{\Delta\tau_{max}^{R2}}\right]^2 + 2\frac{\Delta t_0}{\Delta\tau_{max}^{R1}}\left[\left(\frac{\Delta\tau_{max}^{R1}}{\Delta t_0}\right)^2 - 1\right]^{\frac{1}{2}}\right)^{\frac{1}{2}} + \left[\frac{\Delta t_0}{\Delta\tau_{max}^{R2}}\right]^2\right]^{\frac{1}{2}} \frac{\Delta t_0}{\Delta\tau_{max}^{R2}} \qquad \text{S7}$$

Where, $\Delta\tau_{max}^{R1} = \left[1 + \left(1 + 2\frac{L_N}{L_d}\right)^2\right]^{\frac{1}{2}} \frac{\Delta t_0}{\sqrt{2}}$, $\Delta\tau_{max}^{R2} = \left[1 + \left(\left[\frac{5}{2}\left[1 + \left(1 + 2\frac{L_N}{L_d}\right)^2\right] - 1\right]^{\frac{1}{2}} + 5\frac{L_N}{L_d}\right)^2\right]^{\frac{1}{2}} \frac{1}{\sqrt{5}} \Delta t_0$

The quantities and corresponding labels are explicitly explained in the proof section below.

As expected, $F = \sqrt{5}$ as $\frac{L_{nl}}{L_d} \to 0$ since $\Delta\tau_{max}^{R2}$ and $\Delta\tau_{max}^{R1} \to \Delta t_0$ in this limit and to $F = 1$ as $\frac{L_{nl}}{L_d} \to \infty$ as both $\Delta\tau_{max}^{R1}$ and $\Delta\tau_{max}^{R2}$ go to $\infty$. Approximations can be derived from Eq. S7 by neglecting terms such as $\frac{\Delta t_0}{\Delta\tau_{max}^{R1}}, \frac{\Delta t_0}{\Delta\tau_{max}^{R2}}, \left[\left(\frac{\Delta\tau_{max}^{R1}}{\Delta t_0}\right)^2 - 1\right]^{\frac{1}{2}}$ that remain always less than one and using Taylor series approximations for the square roots.

Important for the generality of the results is that no explicit knowledge of the values of $\Delta t_0, \Delta \omega_0, L_N, L_d$ are needed. Given the above dependencies, it can easily be shown that all that is needed as input to calculate $F$, is the ratio $\frac{L_N}{L_d}$ (or equivalently $\frac{R}{2}$, recalling that $R \equiv \frac{L_{nl}}{L_d}$, as shown in the main text).

In order to provide additional insight to the dynamics of the bandwidth increasement factor, $F$, versus $R$ in the alternating dispersion structure, we obtain the $R_p$ and $F_p$ values at every $p^{th}$ ND segment in the numerical alternating fiber example of Fig. 1D (numbered from ND1 to ND11) obtained from the full GNLSE under constant $\beta_2$. A plot of the $F_p$ values (solid blue curve) versus $R_p$ values is shown in Fig. S1.

For comparison, and to show that Eq. S24 is a lower-bound expression for the bandwidth increase in a ND segment, we also plot the calculated $F_p$ values, using Eq. S24, for the $R_p$ values given by the numerical example. The results of the calculation are shown as a dotted trace in Fig. S1. It can be seen that Eq. S24 displays the same type of dependence vs. $R$, while remaining below the results from the numerical calculation.

Also shown in the numerical example of Fig. S1 is that $F_p$ decreases and $R_p$ increases for increasing ND segment numbers (increasing $p$) in the sign-alternating dispersion structure as described in the main text.

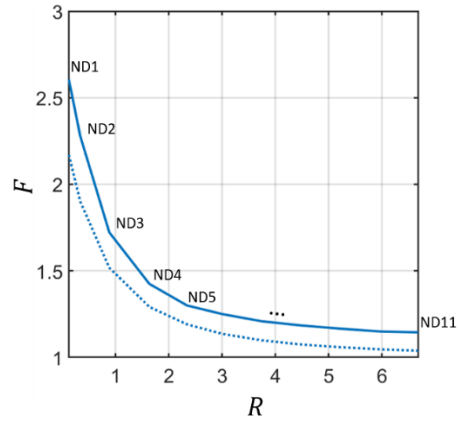

Fig. S1. Solid trace: bandwidth increasement factor, $F$ versus $R$ of the numerical example. The specific ND segment belonging to an $(F, R)$ pair is labelled on the graph. Dotted trace: lower bound calculation according to Eq. S24. The latter is indeed always lower than the numerical data, but has the same type of dependence versus $R$.

## *Proof of Eq. S7*
## *Lower Bound Calculation in Region 1*

To start the lower bound calculation proof, we first derive its value at the end of region $z \in [0, L_N]$, labelled as region 1. By defining regions as multiples of $L_N$, simple expressions emerge due to the normalization with respect to $L_N$ in the expression for spectral propagation. For this proof, we label the pulse, consisting of a transform limited input pulse into region 1 (indexed R1) undergoing full SPM and dispersion as the "actual pulse", labelled $p_1$.

If we construct an input pulse that has
1) the same[1] bandwidth as $p_1$ but
2) has a lower peak power and larger duration than the actual pulse $p_1$ throughout region 1,

this would yield a lower bound bandwidth at the end of region 1. Moreover, this pulse would spectrally broaden less even when only considering SPM compared to the actual input pulse undergoing both SPM and dispersion at the end of region 1. Thus, calculating this lower bound only involves considering SPM with such a pulse. To satisfy both conditions, we take a temporally linearly chirped pulse with duration, $\Delta T$, that is strictly greater than $p_1$ (labeled as $\Delta \tau_{max}^{R1}$) and with matching bandwidth and pulse energy, at the input. The corresponding $L_N$ for this pulse is labelled as $L_N^{R1}$.

To find $\Delta \tau_{max}^{R1}$, the following procedure, throughout region 1 is carried out:

Firstly, we omit dispersion and propagate $p_1$ in the waveguide until $z = L_N$. Here, due to SPM we obtain a temporally linearly chirped Gaussian pulse (labelled as $p_2$), with the same duration as $p_1$. Since $p_2$ is obtained where SPM is maximized (due to omitted dispersion), it has an upper bound bandwidth, given by Eq. S6 at $z = L_N$, that cannot be obtained in reality.

---
[1] Less works too, but then the lower bound estimate would be unnecessarily strict

To account for dispersion as the next step, we use $p_2$ as the input condition in region 1 instead of the original $p_1$. Accounting only for dispersion and omitting, SPM we propagate $p_2$ until the end of region 1 (i.e., $z = L_N$). Here, the duration of the pulse at $z = L_N$ would be the upper bound duration of the actual pulse within $z \in [0, L_N]$. This is the case because the upper bound bandwidth is now used as an initial condition, thus, maximizing the temporal broadening by dispersion (by minimizing $\Delta t_0$, refer to Eq. S2).

Using Eq. S2 with $p_2$ propagated a virtual distance $L_d^{R1}$ to match the input duration of $p_1$ at the start of R1, we arrive at:

$$\Delta \tau_{max}^{R1} = \left[1 + \left(\frac{L_d^{R1} + L_N}{L_d^{R1}}\right)^2\right]^{\frac{1}{2}} \frac{\Delta t_0}{\sqrt{2}} = \left[1 + \left(1 + 2\frac{L_N}{L_d}\right)^2\right]^{\frac{1}{2}} \frac{\Delta t_0}{\sqrt{2}} \qquad \text{S8}$$

with $L_d^{R1} = \frac{L_d}{2} = \frac{\Delta t_0^2}{2|\beta_2|}$, $L_N = \frac{1}{2\gamma P_o} = \frac{L_{nl}}{2}$,

where $P_o$ and $\Delta t_0$ are the peak power and pulse duration of the original pulse $p_1$ at $z = 0$. The above is derived by obtaining the spectral extent at $z = L_N$ in the absence of dispersion as

$$\Delta \omega(L_N) = \left[1 + \left(\frac{L_N}{L_N}\right)^2\right]^{\frac{1}{2}} \Delta \omega_0 = \sqrt{2}\Delta \omega_0 \qquad . \qquad \text{S9}$$

And therefore, the transform limit of this Gaussian pulse, $\Delta \tau_o^{p_2}$, is:

$$\Delta \tau_o^{p_2} \propto \frac{1}{\omega_p} = \frac{1}{\sqrt{2}\Delta \omega_0} = \frac{\Delta t_0}{\sqrt{2}} \qquad . \qquad \text{S10}$$

This is then substituted for the input transform limited temporal duration in Eq. S8. Likewise, the $L_d$ that is applicable for Eq. S8 (labelled as $L_d^{R1}$) is calculated using this input duration, yielding the expression of $L_d^{R1}$ given above.

Now that we have $\Delta \tau_{max}^{R1}$, to find the spectral bandwidth for such a temporally chirped pulse, it can be shown that Eq. S6 becomes:

$$\Delta \omega(z) = \left[1 + \left(\frac{z + z_{on}}{L_N^{R1}}\right)^2\right]^{\frac{1}{2}} \frac{\Delta t_0}{\Delta T} \Delta \omega_0 \qquad \text{S11}$$

$$L_N^{R1} = \frac{1}{2\gamma \frac{\Delta t_0}{\Delta T} P_o} = L_N \frac{\Delta T}{\Delta t_0} \qquad . \qquad \text{S12}$$

Eq. S11 is derived by noting that a chirped pulse is identical to a transform limited pulse of bandwidth $\frac{\Delta t_0}{\Delta T}\Delta \omega_0$ that traversed a virtual distance $z_{on}$ in the waveguide, under only SPM, such that when it enters region 1 it is linearly temporally chirped. $z_{on}$ is found such that Eq. S11 satisfies the input condition 1): At $z = 0$, $\Delta \omega = \Delta \omega_0$, then,

$$z_{on} = \left[\left(\frac{\Delta T}{\Delta t_0}\right)^2 - 1\right]^{\frac{1}{2}} L_N^{R1} \qquad , \qquad \text{S13}$$

since a lower bound spectral broadening is guaranteed using the above pulse choice, then applying Eq. S11 at $z = L_N$ gives the lower bound bandwidth as

$$\Delta \omega^{R1} = \left[1 + \left(\frac{L_N + z_{on}}{L_N^{R1}}\right)^2\right]^{\frac{1}{2}} \frac{\Delta t_0}{\Delta \tau_{max}^{R1}} \Delta \omega_0 = \left[1 + \left(\frac{\Delta t_0}{\Delta \tau_{max}^{R1}} + \left[\left(\frac{\Delta \tau_{max}^{R1}}{\Delta t_0}\right)^2 - 1\right]^{\frac{1}{2}}\right)^2\right]^{\frac{1}{2}} \frac{\Delta t_0}{\Delta \tau_{max}^{R1}} \Delta \omega_0 , \qquad \text{S14}$$

where $L_N^{R1} = \frac{1}{2\gamma \frac{\Delta t_0}{\Delta \tau_{max}^{R1}} P_o} = L_N \frac{\Delta \tau_{max}^{R1}}{\Delta t_0}$, $z_{on} = \left[\left(\frac{\Delta \tau_{max}^{R1}}{\Delta t_0}\right)^2 - 1\right]^{\frac{1}{2}} L_N^{R1}$ .

Using Eq. S14 we obtain a strict lower bound spectral bandwidth ratio (labelled as $LBR^{R1}$) between the input spectrum and the spectrum at $z = L_N$ as

$$LBR^{R1} = \frac{\Delta \omega^{R1}}{\Delta \omega_0} = \left[1 + \left(\frac{\Delta t_0}{\Delta \tau_{max}^{R1}} + \left[\left(\frac{\Delta \tau_{max}^{R1}}{\Delta t_0}\right)^2 - 1\right]^{\frac{1}{2}}\right)^2\right]^{\frac{1}{2}} \frac{\Delta t_0}{\Delta \tau_{max}^{R1}} \qquad . \qquad \text{S15}$$

As the next step, at the end of region 1, as input for a next region, we assume that there is a temporally chirped pulse of temporal duration $\Delta\tau_{max}^{R1}$ and bandwidth $\Delta\omega_p{}^{R1}$. Such assumption warrants that an actual pulse will strictly have a broader spectral bandwidth and a shorter temporal duration. Thus, this assumption will enable us to derive the lower bound spectral bandwidth and lower bound broadening factor, $LBR^{R2}$, encompassing also the next region of propagation, $z \in (L_N, 2L_N]$, labelled as region 2. This factor is the bandwidth increasement factor $F$ used in the main text.

## *Lower Bound Calculation in Region 2 Yielding F*

We now continue this analysis for region 2 (labelled R2, $z \in (L_N, 2L_N]$) to obtain a more relevant $LBR$, at the end of region 2, as this encompasses bandwidth generation and dispersion across the full nonlinear length of the generating segment ($L_{nl}$). The procedure for region 2 mirrors that for region 1, with the same conditions 1), 2) apply here too.

As in region 1, we first start by finding an unattainable upper bound pulse duration, labelled as $\Delta\tau_{max}^{R2}$. In order to do this, the upper bound spectral bandwidth is obtained at the end of region 2 by omitting dispersion:

$$\Delta\omega(2L_N) = \left[1 + \left(\frac{2L_N}{L_N}\right)^2\right]^{\frac{1}{2}} \Delta\omega_0 = \sqrt{5}\Delta\omega_0 \qquad \text{S16}$$

We now construct a temporally linearly chirped pulse (labelled $p_3$) matching at least the same temporal duration of the actual pulse at the start of region 2, with this maximal bandwidth. Thus, at the end of region 2 the temporal duration of $p_3$ would be $\Delta\tau_{max}^{R2}$.

However, we do not know the actual pulse duration at the start of region 2. To resolve this, we need to use as input condition into region 2 a pulse duration for $p_3$ that is known to be greater than the actual pulse duration, $p_1$ at this point. Thus, we use $\Delta\tau_{max}^{R1}$ for the input pulse duration of $p_3$.

To find $\Delta\tau_{max}^{R2}$ we first obtain the expression for the temporal duration of $p_3$. This is given as:

$$\Delta t(z_{R2}) = \left[1 + \left(\left(\frac{\Delta\omega(2L_N)}{\Delta\omega_0}\right)^2 \frac{(z_{od}+z_{R2})}{L_d}\right)^2\right]^{\frac{1}{2}} \frac{\Delta\omega_0}{\Delta\omega(2L_N)} \Delta t_0 \qquad , \qquad \text{S17}$$

where $z_{od}$ is the virtual distance needed to match input pulse durations of $p_1, p_3$, $z_{R2} = z - L_N$. This is derived using the corresponding dispersion length $L_d{}^{R2} = \left(\frac{\Delta\omega_0}{\Delta\omega(2L_N)}\right)^2 L_d$. At the start of region 2 for Eq. S17 to match the initial conditions,

$$\Delta\tau_{max}^{R1} = \left[1 + \left(\frac{5}{L_d}(z_{od})\right)^2\right]^{\frac{1}{2}} \frac{1}{\sqrt{5}}\Delta t_0 \qquad . \qquad \text{S18}$$

Rearranging for $z_{od}$ we obtain:

$$z_{od} = \left[5\left(\frac{\Delta\tau_{max}^{R1}}{\Delta t_0}\right)^2 - 1\right]^{\frac{1}{2}} \frac{L_d}{5} \qquad . \qquad \text{S20}$$

$\Delta\tau_{max}^{R2}$ is found by substituting Eq. S120 and $z = 2L_N$ into Eq. S17:

$$\Delta\tau_{max}^{R2} = \left[1 + \left(\left[5\left(\frac{\Delta\tau_{max}^{R1}}{\Delta t_0}\right)^2 - 1\right]^{\frac{1}{2}} + 5\frac{L_N}{L_d}\right)^2\right]^{\frac{1}{2}} \frac{1}{\sqrt{5}}\Delta t_0 = \left[1 + \left(\left[\frac{5}{2}\left[1 + \left(1 + 2\frac{L_N}{L_d}\right)^2\right] - 1\right]^{\frac{1}{2}} + 5\frac{L_N}{L_d}\right)^2\right]^{\frac{1}{2}} \frac{1}{\sqrt{5}}\Delta t_0$$
$$\text{S21}$$

Once this upper bound pulse duration, $\Delta\tau_{max}^{R2}$, is obtained, the procedure follows exactly the procedure in region 1 to calculate the lower bound spectral bandwidth at $z = 2L_N$, but with $\Delta\tau_{max}^{R2}$ substituted for $\Delta\tau_{max}^{R1}$ and $L_N{}^{R2} = \frac{\Delta\tau_{max}^{R2}}{\Delta t_0} L_N$ substituted for $L_N{}^{R1}$. The lower bound spectral bandwidth at $z = 2L_N$ then becomes:

$$\Delta\omega^{R2} = \left[1 + \left(\frac{L_N+z_{on2}}{L_N{}^{R2}}\right)^2\right]^{\frac{1}{2}} \frac{\Delta t_0}{\Delta\tau_{max}^{R2}} \Delta\omega_0 \qquad . \qquad \text{S22}$$

To solve for the term $z_{on2}$ in Eq. S22, we must use an initial spectral bandwidth into region 2 which must be less than or equal to the actual spectral bandwidth of the pulse going into region 2, to still obtain a strictly lower bound bandwidth at the end of region 2. We could for example use $\Delta\omega_0$ here. However, to obtain a more realistic lower bound at the end of region 2, we use instead $\Delta\omega^{R1}$, because this is strictly lower than the actual pulse bandwidth input into region 2 but still greater than $\Delta\omega_0$. Then, by rearranging Eq. S22 for $z_{on2}$ and using the expression for $\Delta\omega^{R1}$ (Eq. S14) we obtain:

$$z_{on2} = \left[\left(\frac{\Delta\tau_{max}^{R2}}{\Delta t_0}\frac{\Delta\omega^{R1}}{\Delta\omega_0}\right)^2 - 1\right]^{\frac{1}{2}} L_N^{R2} = \left[\left(\frac{\Delta\tau_{max}^{R2}}{\Delta\tau_{max}^{R1}}\left[1 + \left(\frac{\Delta t_0}{\Delta\tau_{max}^{R1}} + \left[\left(\frac{\Delta\tau_{max}^{R1}}{\Delta t_0}\right)^2 - 1\right]^{\frac{1}{2}}\right)^2\right]^{\frac{1}{2}}\right)^2 - 1\right]^{\frac{1}{2}} L_N^{R2} \quad . \quad S23$$

Substituting, Eq. S23 into Eq. S22 yields:

$$\Delta\omega^{R2} = \left[1 + \left(\frac{\Delta t_0}{\Delta\tau_{max}^{R2}} + \left[\left(\frac{\Delta\tau_{max}^{R2}}{\Delta\tau_{max}^{R1}}\left[1 + \left(\frac{\Delta t_0}{\Delta\tau_{max}^{R1}} + \left[\left(\frac{\Delta\tau_{max}^{R1}}{\Delta t_0}\right)^2 - 1\right]^{\frac{1}{2}}\right)^2\right]^{\frac{1}{2}}\right)^2 - 1\right]^{\frac{1}{2}}\right)^2\right]^{\frac{1}{2}} \frac{\Delta t_0}{\Delta\tau_{max}^{R2}}\Delta\omega_0 \quad , \quad S24$$

where we have used $\frac{L_N}{L_N^{R2}} = \frac{\Delta t_0}{\Delta\tau_{max}^{R2}}$, given from the expression of $L_N^{R2}$ shown above. $\Delta\omega^{R2}$ is then the bandwidth at the end of region 2 which, by the above derivation, is guaranteed to be a strict lower bound bandwidth.

Dividing Eq. S23 by the input bandwidth into the segment, $\Delta\omega_0$, we obtain a strict lower bound for the bandwidth ratio (now labelled as $LBR^{R2}$) between the input bandwidth and output bandwidth at the end of region 2, i.e., corresponding to $z = 2L_N = L_{nl}$. This is the bandwidth increasement factor for the segment, $F$, as defined in the paper for segment length $z = L_{nl}$ and algebraically simplifies to Eq. S7. Ending the proof.

$F = LBR^{R2}$ in terms of $LBR^{R1}$ yields a simpler expression namely:

$$F = LBR^{R2} = \left[1 + \left(\frac{\Delta t_0}{\Delta\tau_{max}^{R2}} + \left[\left(\frac{\Delta\tau_{max}^{R2}}{\Delta t_0}LBR^{R1}\right)^2 - 1\right]^{\frac{1}{2}}\right)^2\right]^{\frac{1}{2}} \frac{\Delta t_0}{\Delta\tau_{max}^{R2}} \quad . \quad S25$$

Finally, we note that the analysis of minimum spectral broadening in the considered segment can be further refined, and somewhat larger values for $F$ can be obtained, by considering additional propagation regions of length $L_N$ (e.g., $z \in (2L_N, 3L_N]$) in the identical way as done for region 2. We find the following recursion relationship for the $LBR$ at the end of region $n + 1$ ($z = (n + 1)L_N$) in terms of the $LBR$ at the end of region $n$ ($z = (n)L_N$):

$$LBR^{Rn+1} = \left[1 + \left(\frac{\Delta t_0}{\Delta\tau_{max}^{Rn+1}} + \left[\left(\frac{\Delta\tau_{max}^{Rn+1}}{\Delta t_0}LBR^{Rn}\right)^2 - 1\right]^{\frac{1}{2}}\right)^2\right]^{\frac{1}{2}} \frac{\Delta t_0}{\Delta\tau_{max}^{Rn+1}} \quad , \quad S25b$$

where, $\Delta\tau_{max}^{Rn+1}$ is given by :

$$\Delta\tau_{max}^{Rn+1} = \left[1 + \left(\left[((n+1)^2 + 1)\left(\frac{\Delta\tau_{max}^{Rn}}{\Delta t_0}\right)^2 - 1\right]^{\frac{1}{2}} + ((n+1)^2 + 1)\frac{L_N}{L_d}\right)^2\right]^{\frac{1}{2}} \frac{1}{\sqrt{((n+1)^2 + 1)}} \Delta t_0 \quad .$$

## Derivation of Main Text Eq. 2

Eq. 1 can be approximated by an exponential function shown in Eq. 2, for segments whose $R = \frac{L_{nl}}{L_d} \leq 1$. As a first approximation for the base of this exponential function, in the limit ($R \ll 1$), we omit dispersive effects deriving $F_p$, by substituting $z = L_{nl} = 2L_N$ into Eq. S6 to obtain

$$\Delta\omega_n(z = 2L_N) = [5]^{\frac{1}{2}}\Delta\omega_{n-1} \quad . \quad S26$$

Tracing this to the first generating segment, we obtain: $\Delta\omega_n(z = 2L_N) = [5]^{\frac{1}{2}(n)}\Delta\omega_0$. I.e., the exponential function of Eq. 2 would then have $F_p = [5]^{\frac{1}{2}}$ as base.

However, even for segments where $R \ll 1$, the R value always increases and $F_p$ reduces, as a function of segment number. Nevertheless, it can be shown that $F_p$ is sufficiently slowly decreasing such that Eq. 1 can still be approximated by an exponential function with some average base (i.e., Eq. 2). The best fitting exponential function within this range (i.e., for Eq. 2), would then be constructed by taking as base the geometrical mean of the segment $F_p$ values [1].

We now explicitly prove that the $R$ range, where Eq. 1 can be approximated by Eq. 2 is $R \leq 1$, because $F_p$ reduces sufficiently slowly as a function of segment numbers located in this range. In order to do this, we approximate Eq. 1 as an exponential function when each additional segment $F_p$ in the product is reduced sufficiently slowly such that the bandwidth buildup due to that additional segment is not linear. We correspondingly find the range of $F_p$ values where the growth is not linear and where it may be approximated as exponential. This approximation works within the degree of accuracy we accept for a general understanding of our approach. Further, the validity of this approximation is justified as it can be shown that the transitory region between exponential spectral increase and linear occupies a small fraction of this range. Finally, the corresponding $R$ range matching this interval of $F_p$ values is then the range of validity of Eq. 2.

Proceeding, we prove by induction. To start, if the spectral increase is linear for two neighboring generating segments ($F_1, F_2$, labels the $F_p$ of first, second segment), the following can be derived:

$$F_2 = 2 - \frac{1}{F_1} \qquad \qquad \text{S27}$$

Then if the actual $F_2$ (labelled as $F_2^{(actual)}$) satisfies $F_2^{(actual)} < F_2$, the growth would be greater than linear across these two segments, which we then assume in our approximation as exponential.

As the maximal value $F_1$ can be $\sqrt{5}$, $F_2$ would be maximized as well. This has the effect that the corresponding $R$ for $F_2$ would be at the minimum possible. Eq. S28 gives $F_2 = 1.57$. From this, the corresponding $R = 0.8$. We take this value as our $R$ range here as it is the most conservative. Therefore, for $R < 0.8$, $F_2^{(actual)} < F_2$ and the exponential range is then determined as valid for $R \leq 0.8$. In reality, since $F_1 < \sqrt{5}$, we take this as justification to round up the range of $R$ comprising approximately exponential growth as $R \leq 1$.

For the next pair of neighboring segments, we obtain $F_3 = 2 - \frac{1}{F_2^{(actual)}}$ for linear growth, since $F_n$ monotonically decreases as a function of $n$, $F_2^{(actual)} < F_1$. Then, $F_3 < F_2$ and the corresponding $R$ for $F_3$ to satisfy linear growth would be even greater than $R = 1$. Thus, if $R_3 < 1$, exponential growth is maintained between generating segment 2 and 3. The above can be extended for the additional neighboring generating segment pairs, i.e., in general, $F_n = 2 - \frac{1}{F_{n-1}^{(actual)}}$, is not satisfied for $R_n < 1$. Consequently, for all segments where $R < 1$, Eq. 1 can be approximated by an exponential function, i.e., Eq. 2.

Having determined the range of $R$ where Eq. 2 is approximately valid, we conclude this section by addressing how the validity of Eq. 2 increases just by decreasing $R_1$ (i.e., by increasing pump power, or by manipulating the nonlinear coefficient, $\gamma$ of materials). This result shows that increasing input peak power into the alternating structure, would increase the validity of the approximation of Eq. 1 to reducing to Eq. 2. To prove this, first, since $R_2 = F_1 R_1$ (through $R \propto \frac{1}{\Delta t_0}$) and $F_1$ saturates or is slowly growing when $R_1 < 1$, as $R_1 \to 0$ the interval between $R_2$ and $R_1$ reduces. In general, since

$$R_n = F_{n-1} R_{n-1} \qquad , \qquad \qquad \text{S28}$$

the interval between $R_3$ and $R_2$ would then reduce and so forth. Thus, the $F$ values of more segment lie within $R \leq 1$ as $R_1 \to 0$ and there are increasingly more segments per propagation distance. Consequently, the $R$ intervals between adjacent segment $F$'s reduce, strongly satisfying the conditions for exponential build up, namely that the $F$ values vary slowly and the approximation of Eq. 1 by Eq. 2 becomes better.

In general, when $R < 1$, the $R$ interval spacing of segments here are smaller, and thus the region has more segments per a fixed interval than when $R > 1$, where the R interval spacing of segments are larger due to $R_{n-1}$ being greater than unity (in Eq. S28). In fact, since the transitory R region where exponential growth becomes linear growth is located at $R \approx 1$, the amount of segments in this region is therefore less than in the exponential case. Thus, the dynamics are dominated by exponential buildup and then linear build up throughout the structure. The figure for the lower bound estimate (Fig. 1D) is shown in a log-log plot to capture this interval structure of segment $F$'s and $R$'s.

## Supplementary Section 3

### Derivation of Main text Eq. 3

To begin the derivation of main text Eq. 3, we note that in dispersion dominated systems ($L_{d,n} \ll L_{nl,n}$), simple closed form expressions describing the spectral increase can be found. They describe in a lower bound fashion the expected spectral increase, and indicate that the spectral increase is linear with pulse energy, and linear across increasing ND segment numbers in the alternating structure.

To begin deriving these expressions, the first approximation we use is that the maximal instantaneous frequency endpoints (blue and red side) that occur in the temporal profile of the pulse are the endpoints of the frequency bandwidth. This is always true for Gaussian pulses, who have propagated a distance equal to the dispersion length, in the dispersive limit $L_{d,n} \ll L_{nl,n}$. Also true for $z \geq L_{d,n}$ is that the intensity gradient at these endpoints integrated over the propagation distance is directly proportional to the additive bandwidth increase. For distances less than the dispersion length, we assume that there is zero spectral generation. This is a strict criterion which enables us to warrant that the spectral bandwidth generation in the segment is greater than or equal to the right side term in Eq. 3. Thus, from the above approximations we obtain

$$\frac{\partial \Delta \omega_b}{\partial z} = \gamma \left[ \left| \frac{dP}{dt} \right|_{max} \right], \frac{\partial \Delta \omega_r}{\partial z} = \gamma \left[ \left| \frac{dP}{dt} \right|_{min} \right] . \qquad \text{S29}$$

Here, $P$ is the peak power, $\Delta \omega_b$ is the blue side angular frequency bandwidth addition, and $\Delta \omega_r$ the red side angular frequency bandwidth addition.

We have used the expression for the temporal evolution of a transform limited Gaussian pulse undergoing only ND 2$^{nd}$ order dispersion to represent the actual temporal evolution here with SPM. This is justified when the amount of spectral generation is small compared to the initial bandwidth of the pulse, which arises when $L_{d,n} \ll L_{nl,n}$. From Eq. S2, the expression for Gaussian peak power as it broadens, and the expression for the maximum absolute values for derivatives of a Gaussian function, make the temporal intensity derivatives of Eq. S29 become:

$$\left| \frac{dP}{dt}_{max \, or \, min} \right| = g_o \frac{E}{\left[1 + \left(\frac{z^2}{L_D^2}\right)\right] \tau_{po}^2} . \qquad \text{S30}$$

Here, $E = \sqrt{\pi} P \Delta t$ is the pulse energy, $g_o = 0.81$ is a constant related to the Gaussian pulse shape. The integral representation of the differential Eq. S28, in the region of interest past one dispersion length is

$$\Delta \omega_{b,r} = \int_{L_D}^{L} \gamma \left[ \left| \frac{dP}{dt}_{max.min} \right| \right] dz . \qquad \text{S31}$$

We proceed by substituting Eq. S30 into Eq. S31 and solving the trigonometric integral that results to obtain:

$$\Delta \omega_{b,r} = g_o \gamma_2 \frac{E}{|\beta_2|} \left[ \tan^{-1} \frac{L}{L_D} - \frac{\pi}{4} \right] . \qquad \text{S32}$$

Having found the bandwidth additions $\Delta \omega_r$ and $\Delta \omega_b$ for each generating segment, the total bandwidth after $n$ generating segments is then given as:

$$\Delta \omega_n = \Delta \omega_r + \Delta \omega_b + \Delta \omega_0 = n\pi \left[ g_o \gamma_2 \frac{E}{2|\beta_2|} \right] + \Delta \omega_0 \qquad \text{S33}$$

Its frequency representation would then be $\Delta \nu_n = n \left[ g_o \gamma_2 \frac{E}{4|\beta_2|} \right] + \Delta \nu_0$. \qquad S34

### Power Scaling Laws Derivation

When $P_O$ is in the low power limit, such that $R_1 \gg 1$, then at the first segment, Eq. 3 applies with $n = 1$ and therefore, the growth is linear with input peak power, $P_O$ (but upper bounded by a function $\propto \sqrt{P_O}$) .

It is always true that $R_{n+1} > R_n$ due to the increasing bandwidth across the ND segments of the chain, so if $R_1 \gg 1$, also $R_n \gg 1$ is fulfilled. Under this limit, accordingly main text Eq. 3 (Eq. S33) applies for all segments and they will scale linearly with $P_O$ (since Eq. 3 scales linearly with $P_O$ through its linear scaling with pulse energy).

Thus, $\Delta\omega_n = n\pi \left[g_o \gamma_2 \frac{E}{2|\beta_2|}\right] \propto nP_O$. While for the uniform dispersion waveguide, seen as one segment, $\Delta\omega \propto P_O$ in this limit.

As $P_O$ increases, the first segment $R^{(1)}$ becomes smaller than unity at some input peak power. It can be shown that when $R < 10^{-1}$, $F_p$ at least scales close to $\sqrt{P_O}$ here as well. Thus, we use $F_p \propto \sqrt{P_O}$ in Eq. 2 and then the scaling here would be: $\Delta\omega_n = A\sqrt{P_O} + B(n-1)P_O$ (if only the first segment $R < 1$), where, $A > B$ and are scaling constants. As $P_O$ increases further, the approximate exponential regime is valid for an increasing number of segments in the structure, since $R < 1$ is satisfied for more segments (e.g., $R_1, R_2, R_3, etc.. < 1$). It can be proved from the lower bound calculation that the geometrical mean (see discussion in Eq. 2 derivation section) for the exponential function describing the spectral build-up across these segments would have base $F_p \propto \sqrt{P_O}$. This continues until the bandwidth grows to high values and the linear regime applies for segments whose $R > 1$.

To quantify the transition from exponential to linear growth, segment number $n_t$ labels the first segment where $R > 1$. While growth for segments in the transitory region is still somewhat greater than linear (e.g. polynomial), for the sake of simplicity we lump these segments together with the linear growth segments, making the power scaling laws we present here a conservative estimate. In this $P_O$ power region then, $\Delta\omega_n = \left(A\sqrt{P_O}\right)^{n_T-1} + B(n - n_T + 1)P_O$. For a more compact expression we replace the equation with constants $A, B$ with a proportionality law and represent this overall as $\Delta\omega_n \propto (c_T\sqrt{P_O})^{n_T-1} + (n - n_T + 1)P_O$, $c_T \equiv \frac{A}{B^{-n_T+1}} > 1$.

We note that when we increase $P_O$ further, at the extreme power limits, the lower bound estimate presented above ($\Delta\omega_n \propto (c_T\sqrt{P_O})^{n_T-1} + (n - n_T + 1)P_O$), must be adjusted. The reason is that at very small $R$ values (i.e., $R < 10^{-2}$), the lower bound estimate would have to be extended past region 2, i.e., $z_p > L_{nl}$, because the stagnation length now is much greater than the nonlinear length. Here, $F_p$ is $\propto \sqrt{\frac{1}{R_p}}$ for all segments in the region $R \ll 1$ (translating to $F_p \propto \sqrt{P_p}$, where, $P_p$ is the peak power entering segment number $p$) as given in main text references. This translates to a fast build up the spectrum and large relative $R$ jumps for subsequent segments (in the region $R \ll 1$). It can be shown through the recursive relationships that govern $R$ and $F_p$ (Eq. S28) that the following power scaling law can be derived here : $\Delta\omega_n = (A)^{n_T-1}P_O + B(n - n_T + 1)P_O = [c_T{}^{n_T-1} + (n - n_T + 1)]P_O$. Where we preserve the conservative scaling by ignoring segments within the $R$ range between $R \ll 1$ and $R > 1$.

To give an overview, the described results are summarized in the following table of power scaling that rules spectral broadening in the dispersion alternating structure, compared to spectral broadening in a conventional structure with uniform dispersion.

Table S3. Overview on spectral bandwidth vs input peak power scaling laws

| Regime | Alternating structure | Uniform dispersion waveguide |
|---|---|---|
| $R_1 > 1$ | $\Delta\omega_n \propto nP_o$ | $\Delta\omega \propto P_o$ |
| $R_1 \leq 1$ | $\Delta\omega_n \propto (c_T\sqrt{P_O})^{n_T-1} + (n - n_T + 1)P_O$ | $\Delta\omega \propto P_o^{\frac{1}{2}}$ |
| $R_1 \ll 1$ | $[c_T{}^{n_T-1} + (n - n_T + 1)]P_O$ | $\Delta\omega \propto P_o^{\frac{1}{2}}$ |

An interesting case arises when the generating segment is the same (i.e. comprises the same material, geometry, with the same length or with a length up to where spectral broadening stagnates, i.e., stops increasing) as the uniform dispersion waveguide being compared to. Then, the spectral bandwidth in the alternating structure will be greater for any power regime, in addition to the steeper scaling.

The above power scaling laws apply where losses and third order dispersion are negligible for both ND and AD segments. Otherwise they are valid over a restricted number of segments in the structure. However, we note that, already at $n_f = 2$, wider bandwidths are achieved than with conventional power scaling laws in uniform media.

# Supplementary Section 4

The full experimental splice losses, lengths and material nonlinear coefficients are listed in this section, under table S4.

Table S4. Segment type, splice losses and length

| Fiber Type & Segment Order | Segment Length (cm) | Splice Loss (dB) | $\gamma$ |
|---|---|---|---|
| **1. Hi1060flex** | 15 | | $5.2\text{E}-3\ W^{-1}m$ |
| **2. SMF28** | 10 | $-0.30$ | $1.1\text{E}-3\ W^{-1}m$ |
| **3. Hi1060flex** | 10 | $-0.63$ | $5.2\text{E}-3\ W^{-1}m$ |
| **4. SMF28** | 8* | $-0.38$ | $1.1\text{E}-3\ W^{-1}m$ |
| **5. Hi1060flex** | 10 | $-0.64$ | $5.2\text{E}-3\ W^{-1}m$ |
| **6. SMF28** | 7 | $-0.40$ | $1.1\text{E}-3\ W^{-1}m$ |
| **7. Hi1060flex** | 15 | $-0.56$ | $5.2\text{E}-3\ W^{-1}m$ |
| **Total** | 75 | $-2.91\ (\approx 49\%)$ | --- |

* modified to 6 cm in the simulation because better agreement to experimental data was found. The deviation of 2 cm from the experimental vale might be due to slight deviations between the calculated and actual dispersion parameter curves or due to the effects of the splice joints on pulse propagation.

# Supplementary Section 5

The waveguide parameters used in the simulation for Fig. 3 in the main text are displayed in table S5.

Table S5 waveguide type, segment length, nonlinear coefficient

| Waveguide Type & Segment Order | Segment Length (mm) | $\gamma$ |
|---|---|---|
| **1. ND** | 3 | $2.235\ W^{-1}m$ |
| **2. AD** | 18.5 | $0.3299\ W^{-1}m$ |
| **3. ND** | 1 | $2.235\ W^{-1}m$ |
| **4. AD** | 7 | $0.3299\ W^{-1}m$ |
| **5. ND** | 0.5 | $2.235\ W^{-1}m$ |
| **6. AD** | 3 | $0.3299\ W^{-1}m$ |

| 7. ND | 0.8 | 2.235 $W^{-1}m$ |
|---|---|---|
| **Total** | 33.8 | --- |